# Superconductor-insulator transitions in infinite-layer nickelates controlled via *operando* monitored reduction

**Heng Wang[a,b,1], Haoliang Huang[a,b,1,\*], Wei Lv[b, 1], Xianfeng Wu[b,1], Guangdi Zhou[a,b,\*], Zihao Nie[b], Yueying Li[b,c], Cui Ding[a, b], Danfeng Li[d,e], Hongtao Yuan[c,\*], Qi-Kun Xue[a,b,f], Zhuoyu Chen[a,b,\*]**

[a]*Quantum Science Center of Guangdong-Hong Kong-Macao Greater Bay Area, Shenzhen 518045, China*
[b]*State Key Laboratory of Quantum Functional Materials, Department of Physics, and Guangdong Basic Research Center of Excellence for Quantum Science, Southern University of Science and Technology, Shenzhen 518055, China*
[c]*National Laboratory of Solid-State Microstructures, College of Engineering and Applied Sciences, Nanjing University, Nanjing, 210008, China.*
[d]*Department of Physics, City University of Hong Kong, Kowloon, Hong Kong SAR 999077, China.*
[e]*Shenzhen Research Institute of City University of Hong Kong, Shenzhen 518057, China.*
[f]*Department of Physics, Tsinghua University, Beijing 100084, China*
[1] *These authors contributed equally to this work.*
*\* Corresponding author.*
*E-mail address: huanghaoliang@quantumsc.cn (H. Huang), zhouguangdi@quantumsc.cn (G. Zhou), htyuan@nju.edu.cn (H. Yuan), chenzhuoyu@sustech.edu.cn (Z. Chen).*

## Abstract

Nickelates represent an emerging class of superconductors that demand innovative approaches for structural and electronic phase modulations. Continuous control over superconductor-insulator transition (SIT) in nickelates remains particularly challenging, hindering both fundamental understanding and potential applications. Here, we demonstrate SIT in infinite-layer nickelate superconductors utilizing multiple techniques, including an *operando* monitored reduction (OMR) method. OMR enables ultrawide-range continuous modulation of the Ni $3d$ orbital electron occupancy from $\sim 3d^7$ to $\sim 3d^9$. The $3d$ occupancy is calibrated through systematic synchrotron X-ray absorption (XAS), combined with scanning transmission electron microscopy (STEM) annular bright field (ABF) analysis of oxygen atoms. SIT is further modulated via ionic liquid gating and magnetic field. Strikingly different from cuprates, our Nernst effect measurements show that pairing initiates at the onset of the resistive drop. The subsequent emergence of the Meissner effect at zero resistance marks the establishment of global phase coherence. Angle-dependent magnetotransport within the transition temperature regime indicates a mixture of two-dimensional (2D) and three-dimensional (3D) superconducting characters, suggesting the observed SIT deviates from the canonical 2D model. Our results provide a unique perspective on the interplay of structural and electronic phase transitions in the infinite-layer nickelates across the oxygen content-magnetic field-temperature parameter space.

## 1. Introduction

For decades, cuprates [1–6] and iron-based compounds [7–13] have been the primary archetype for studying high-transition temperature ($T_C$) superconductivity. The recent advent of nickelate high-$T_c$ superconductors marks a pivotal shift in the field [14–19]. In particular, infinite-layer nickelates share key structural and electronic features with the cuprates [14]. Their similarities and subtle differences offer a critical opportunity to disentangle the complex interplay of factors governing high-$T_c$ superconductivity [18,19].

Precisely controlling the superconductor-insulator transition (SIT) is crucial for understanding the emergence of superconductivity, yet it remains a central challenge in infinite-layer nickelates. The superconducting phase is induced by a topotactic reduction of a perovskite precursor [14,18–26]. This process fundamentally transforms the material by simultaneously altering its crystal and electronic structures. Structurally, it removes apical oxygen to form the square-planar $NiO_2$ layers. Electronically, it can shift the Ni $3d$ orbital occupancy by as many as two electrons.

Pathways ranging from soft chemical reduction using $CaH_2$ [14,18,19,23–29] to *in situ* hydrogen [30–32] or metal reductants [33] have been established, but this reduction has largely functioned as a terminal, single-step process, which offers limited access to or fine control over the intermediate states [34]. Consequently, a continuous and systematic investigation of the physics across the SIT has remained elusive, hindering a deeper understanding of the nickelate superconducting mechanism and its potential applications, despite recent research in emerging nickelate superconductors [35,36].

Herein, we introduce a strategy to overcome this challenge: an *operando* monitored reduction (OMR) method. This technique enables continuous tuning across the nickelate SIT with exceptional fine control, allowing us to map the superconducting phase diagram across a wide parameter space. Leveraging this capability, we systematically investigate the phase diagram of nickelate superconductors by tuning not only the oxygen content but also applying ionic liquid gating and magnetic fields. These results provide a comprehensive framework for understanding and manipulating superconductivity in infinite-layer nickelates.

## 2. Experimental methods

The $Nd_{0.8}Sr_{0.2}NiO_3$ thin films were deposited on (001)-oriented $(LaAlO_3)_{0.3}(Sr_2TaAlO_6)_{0.7}$ (LSAT) substrates by a pulse laser deposition technique. The crystal structure and thickness were characterized using X-ray diffraction (XRD) $\omega$-$2\theta$ scan and X-ray reflectivity measurements. The samples were then placed in the self-developed OMR device to conduct topotactic reduction. Electrical transport measurements were performed in a Physical Property Measurement System with a base temperature of 1.8 K and a magnetic field of up to 14 T. For Nernst effect measurements, we applied a setup featuring one heater and two thermometers. The cross-section scanning transmission electron microscopy (STEM) specimens from our films were prepared using an FEI Helios 600i dual-beam FIB/scanning electron microscope. The valence state characterization of thin films was performed at the BL07U beamline of the Shanghai Synchrotron Radiation Facility (SSRF) [37,38] (see the Supplementary material for details).

## 3. Result and discussion

$Nd_{0.8}Sr_{0.2}NiO_2$ thin films with an infinite-layer structure were synthesized through OMR from perovskite $Nd_{0.8}Sr_{0.2}NiO_3$ precursors grown on LSAT substrates. Fig. 1a shows the high-angle annular dark field (HAADF) STEM image of the $Nd_{0.8}Sr_{0.2}NiO_2$ thin film. We performed combined electrical transport, mutual inductance, and Nernst effect measurements on a superconducting $Nd_{0.8}Sr_{0.2}NiO_2$ film. The results are summarized in Fig. 1b. The film exhibits a superconducting transition, characterized by an onset critical temperature $T_{c,onset}$ of ~ 18 K and a zero-resistance temperature $T_{c0}$ of ~ 12 K. The mutual inductance technique was utilized to characterize the Meissner effect, confirming the onset of diamagnetism at a characteristic temperature $T_M$ of ~ 11 K. This provides evidence for the formation of a phase-coherent superconducting state with a global diamagnetic response. Based on the mutual inductance measurements, we derived the temperature dependence of the pairing coherence length and penetration depth from the critical field of 95% $R_n$, as shown in Fig. 1c.

Fig. 1b presents Nernst effect data for $Nd_{0.8}Sr_{0.2}NiO_2$ at this doping level, showing the temperature-dependent Nernst peak and slope. These measurements reveal evidence of vortex-like fluctuations above the onset critical temperature $T_{c,onset}$. This contrasts with cuprate superconductors, where Nernst measurements revealed vortex-like superconducting fluctuations extending to temperatures significantly above $T_C$ (a phenomenon commonly attributed to precursor pairing of Cooper pairs and associated with the pseudogap state [39–42]). The absence of such fluctuations above $T_{c,onset}$ in $Nd_{0.8}Sr_{0.2}NiO_2$ suggests the absence of a pseudogap state at this doping level. This result needs to be verified through complementary experiments such as angle-resolved photoemission spectroscopy and scanning tunneling microscopy/spectroscopy. Samples spanning a broader doping range should be examined to establish how precursor Cooper-pair correlations evolve with doping levels.

Figs. 1d–f present the upper critical field versus temperature ($H_{c2}$-$T$) phase diagrams for an $Nd_{0.8}Sr_{0.2}NiO_2$ thin film under both out-of-plane ($H // c$-axis) and in-plane ($H // ab$-plane) magnetic fields. The critical temperatures $T_c$ were determined from resistive transitions via three criteria: $T_{c, onset}$ (95% $R_n$), $T_{c, mid}$ (50% $R_n$), and $T_{c, zero}$ (10% $R_n$). Quantitative analysis of $H_{c2}$ for both field orientations, particularly at lower temperatures beyond the critical regime, requires consideration of both orbital and paramagnetic limiting effects, as described by the Werthamer-Helfand-Hohenberg (WHH) theory [43]. The in-plane coherence length $\xi_{ab}$, calculated using

$$\xi_{ab} = \left( \frac{\Phi_0}{2\pi H_{c2,\perp}} \right)^{1/2} \tag{1}$$

with $H_{c2,\perp}$ representing the perpendicular upper critical field, is shown in the right panel of Fig. 1c.

To further probe the dimensionality of superconductivity in $Nd_{0.8}Sr_{0.2}NiO_2$, angle-dependent magnetoresistance measurements were performed on a sample mounted on a probe with rotator under a constant magnetic field of 14 T. The critical temperatures $T_{c,10\%}$, $T_{c,50\%}$, and $T_{c,95\%}$ were extracted using the resistance values corresponding to 10%, 50%, and 95% of the linear fit to the normal-state resistance. The resulting $T_c$ versus rotation angles $\theta$ presented in Figs. 1g–i were analyzed using a model combining 2D and 3D superconducting behaviors. Specifically, we employed a weighted average of Tinkham's model (for 2D superconductors) and the anisotropic Ginzburg-Landau (GL) model (for 3D superconductors), represented by the equation [44,45] :

$$T_c(\theta) = T_{c0} + (1-\beta) \times \{ H_0 / (\partial H_c^{H // c} / \partial T) \times [\sin^2(\theta) + \gamma \cos^2(\theta)]^{1/2} \}$$
$$+ \beta \times [(T_c^{H // c} - T_{c0}) \sin(\theta) + (T_c^{H \perp c} - T_{c0}) \cos^2(\theta)] \tag{2}$$

where $\beta$ is a dimensionless weighting parameter ($0 \leq \beta \leq 1$) quantifying the relative contribution of 2D-like behavior ($\beta \to 1$ indicates stronger 2D character); $\gamma$ is the anisotropy parameter ($\gamma = (m_c / m_{ab})$), representing the ratio of effective masses for out-of-plane ($m_c$) and in-plane ($m_{ab}$) electron motion;

$\theta$ is the angle between the applied magnetic field and the sample's $c$-axis (normal to the sample surface), as defined in the inset of Fig. 1d. Our analysis reveals that the sample possesses a mixed 2D and 3D superconducting dimensionality, deviating from canonical models, with the relative weight of the 2D-like component increasing at lower temperatures. The effective dimensionality of superconductivity in infinite-layer nickelates remains unresolved. Unlike cuprates, where a single $CuO_2$ band dominates, coexisting Nd $5d$ electron pockets and Ni $3d_{x^2-y^2}$ holes sustain finite inter-layer hopping, thus conventional anisotropic Fermi-liquid and quasi-2D Tinkham–Lawrence–Doniach descriptions cannot reproduce our data alone. Whether an extended multiband 2D Hubbard or t-J model can capture the same observations remains to be tested; quantitative many-body calculations will be required to settle this issue.

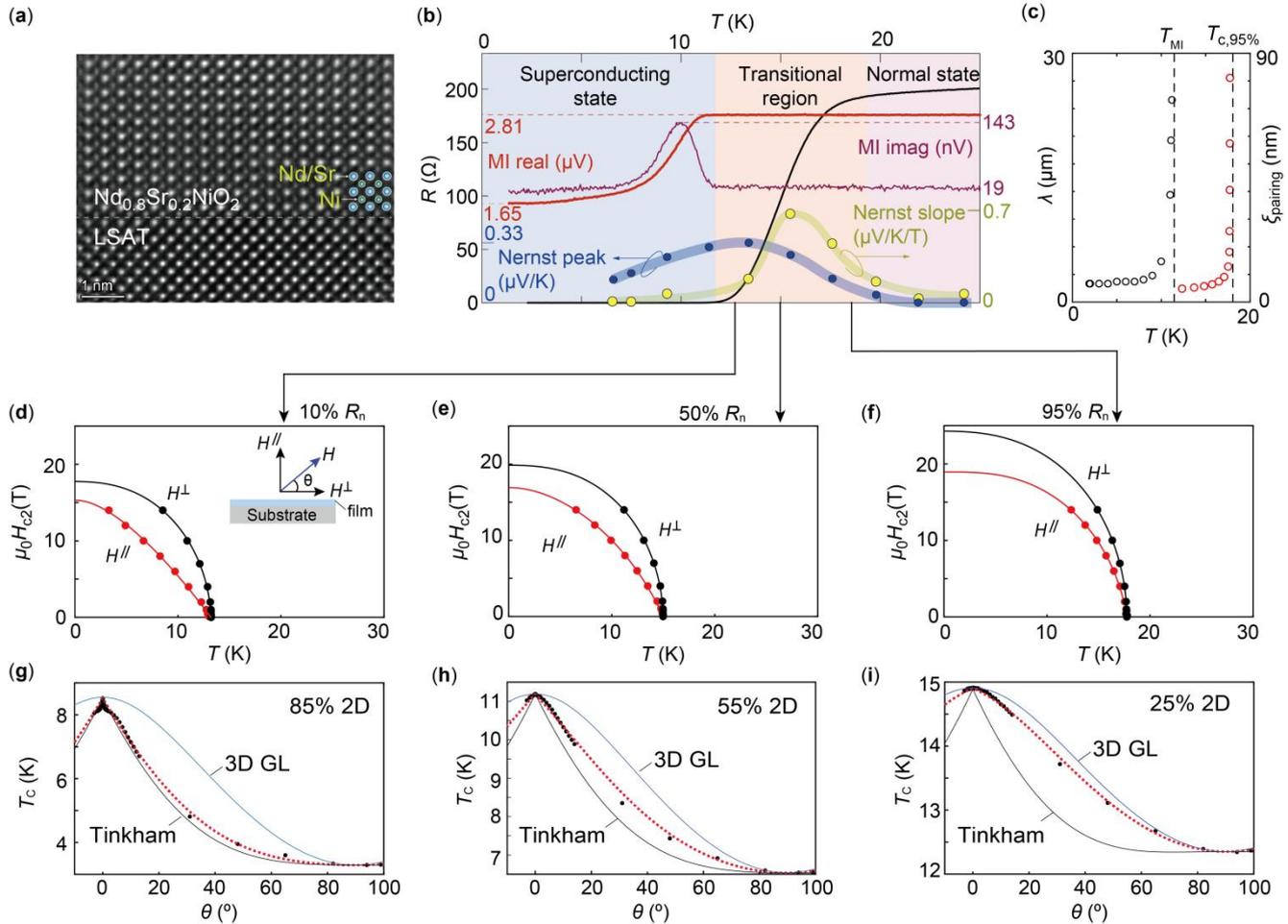

**Fig. 1.** Combined electrical, magnetic and thermal transport measurements of superconducting $Nd_{0.8}Sr_{0.2}NiO_2$ film on LSAT substrate. (a) STEM HAADF image of $Nd_{0.8}Sr_{0.2}NiO_2$ film on (001)-LSAT substrate. (b) Transport measurements: black curve: temperature dependence of electrical resistance. Red and purple curves: mutual inductance measurements. Light blue and yellow: temperature dependence of the Nernst signal. (c) Temperature dependence of the pairing coherence length (derived from the critical field at 95% $R_n$) and penetration depth (obtained from mutual inductance measurements). (d–f) Temperature-dependent upper critical field ($H_{c2}$) for magnetic fields oriented along the c-axis (red lines) and within the a-b plane (black lines). Here, $H_{c2}$ in panels (d–f) is defined as the field strength where the resistance reaches 10%, 50%, and 95% of $R_n$, respectively. $R_n$ denotes the normal-state sheet resistance obtained by a linear extrapolation of the 25–30 K data. (g–i) Angular dependence of the superconducting transition temperature $T_c$ under a magnetic field of 14 T, fitted with a combination of the 2D Tinkham model (black lines) and the 3D Ginzburg-Landau (GL) model (blue lines).

Fig. 2a displays a schematic diagram of the self-developed OMR device. In this setup, the thin film and $CaH_2$ pellet are placed on separate heater sheets. Using this OMR device, the reduction process of the $Nd_{0.8}Sr_{0.2}NiO_3$ film can be tracked in real-time, including the film's resistance, the vacuum chamber pressure, and the sample temperature. Fig. 2b illustrates the temporal evolution of these parameters during the reduction of a typical $Nd_{0.8}Sr_{0.2}NiO_3$ thin film. Initially, Ni exhibits 6.8 nominally occupied electrons in the $3d$ orbitals of the precursor perovskite phase. As the temperature rises, $CaH_2$ is decomposed, increasing the hydrogen concentration within the chamber. Consequently, the film progressively loses oxygen and undergoes reduction, leading to a sharp rise in resistance until the film transforms into the $Nd_{0.8}Sr_{0.2}NiO_{2.5}$ phase, where the Ni $3d$ orbital occupancy increases to 7.8. When the temperature is held at 400 °C and the reduction continues, the film resistance starts to decrease while the chamber pressure gradually increases. Experimentally, the resistance minimum coincides with the highest metallic slope and best superconducting transition. We therefore use this minimum as a reproducible marker for the optimal approach to the infinite-layer phase under our OMR protocol. Prolonged reduction leads to an over-reduced state, causing the resistance to rise slightly again. In the infinite-layer $Nd_{0.8}Sr_{0.2}NiO_{2.0}$ phase, the Ni $3d$ orbital occupancy is 8.8.

XRD and X-ray absorption spectroscopy (XAS) measurements were conducted on samples at six distinct stages of the reduction process, including the initial precursor perovskite phase, intermediate phase, and final over-reduced state. As anticipated, the diffraction peaks of the thin film shifted to higher angles as the reduction progressed, with the out-of-plane lattice parameters decreasing until the infinite-layer phase was fully formed, as illustrated in Fig. 2c. In the over-reduced phase, the film's peak intensity weakened considerably, indicating structural disruption of the infinite-layer phase. Fig. 2d displays the XAS spectra of the Ni L absorption edge of the samples with different reduction stages. A significant change in the $L_3$ absorption peak and a marked increase in the peak intensity at 850.7 eV were observed as the oxygen content decreased. By fitting the Ni $L_3$ absorption peaks of these samples and analysing the intensity differences between the peaks at positions 850.7 and 851.5 eV, we calibrated the occupancy of Ni $3d$ orbital electrons in these samples, as shown in Fig. 2b. These results evident that the control of the Ni valence states over an ultrawide range (from $\sim 3d^7$ to $\sim 3d^9$) can potentially be achieved using OMR while simultaneously maintaining ultrahigh precision.

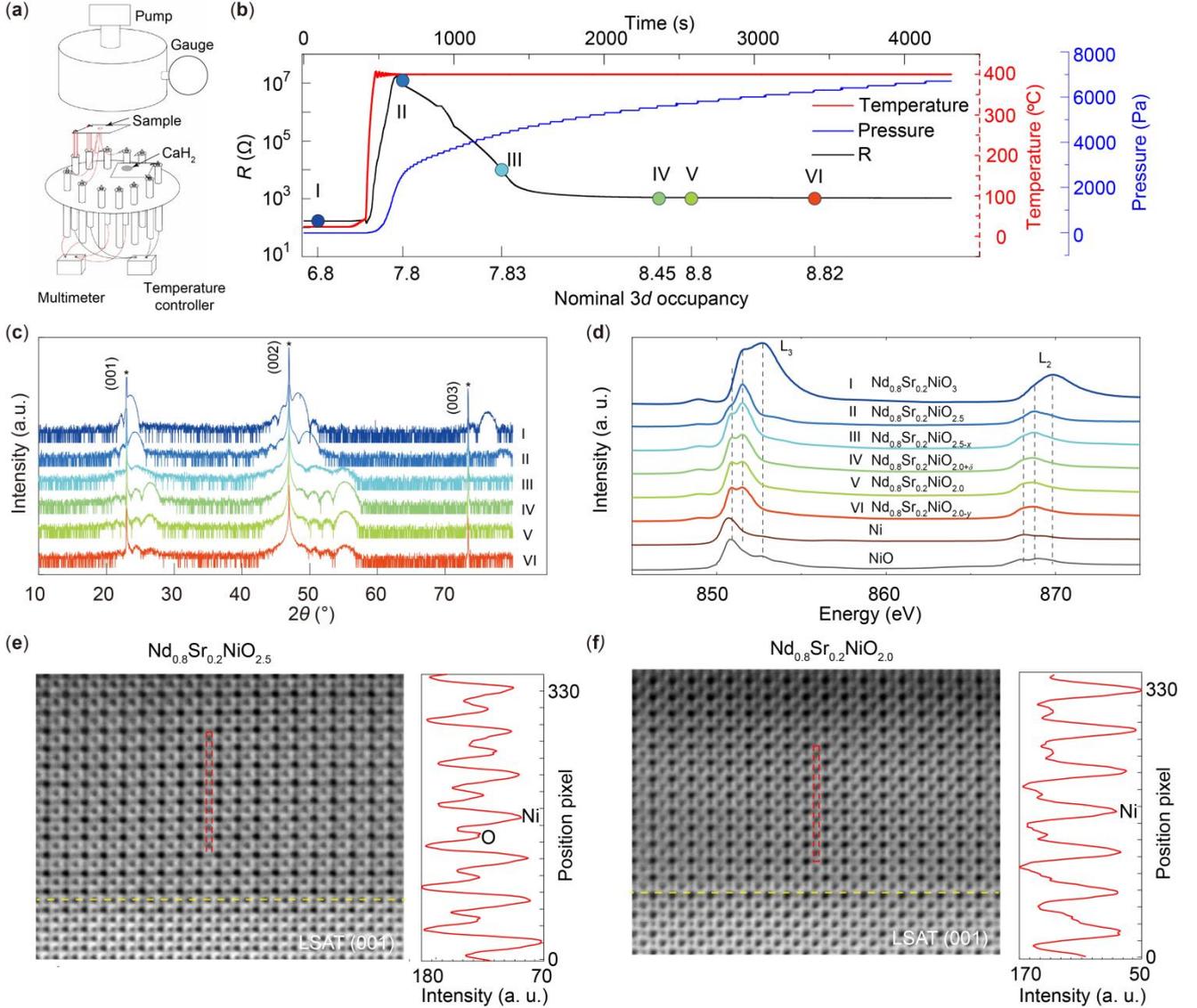

**Fig. 2.** The OMR process from $Nd_{0.8}Sr_{0.2}NiO_3$ thin film to superconducting $Nd_{0.8}Sr_{0.2}NiO_2$ thin film. (a) Schematic diagram of OMR device for the reduction process. (b) The *operando* temperature-pressure-resistance trajectory illustrates the approximate set-points used for the six reduction stages, as recorded from a representative sample; each stage was performed on a separate sister sample The horizontal coordinate at the bottom is the Ni $3d$ occupancy calibrated based on XAS Ni $L$ edges. (c) The comparison of XRD results between the selected six different reduction stages. (d) The Ni $L$-edge XAS spectra of the different reduction stages compared to the standard Ni and NiO. The analysis of oxygen in the ABF image for (e) $Nd_{0.8}Sr_{0.2}NiO_{2.5}$ and (f) $Nd_{0.8}Sr_{0.2}NiO_{2.0}$. Right panels: the intensity of the selected vertical box marked by red dash lines. In the right panel of (f), the residue intensity at the oxygen positions is likely from the tail intensity of adjacent Nd atoms. The yellow dash lines indicate the interface between the film and the substrate.

Oxygen content can be quantified by analyzing the oxygen intensity in annular bright field (ABF) images. Fig. 2e and f display ABF images of the $Nd_{0.8}Sr_{0.2}NiO_{2.5}$ and $Nd_{0.8}Sr_{0.2}NiO_{2.0}$ samples. By comparing the intensity of Ni-O contours in both vertical and horizontal directions, it is evident that apical oxygen content decreases significantly during reduction, becoming nearly undetectable in the infinite-layer phase (Fig. 2f). Furthermore, these ABF images suggest that apical oxygen is preferentially lost during reduction. Quantification of absolute oxygen content from STEM is hindered by a strong background signal, so the data are intended to reveal the relative evolution of apical versus in-plane oxygen rather than precise stoichiometries.

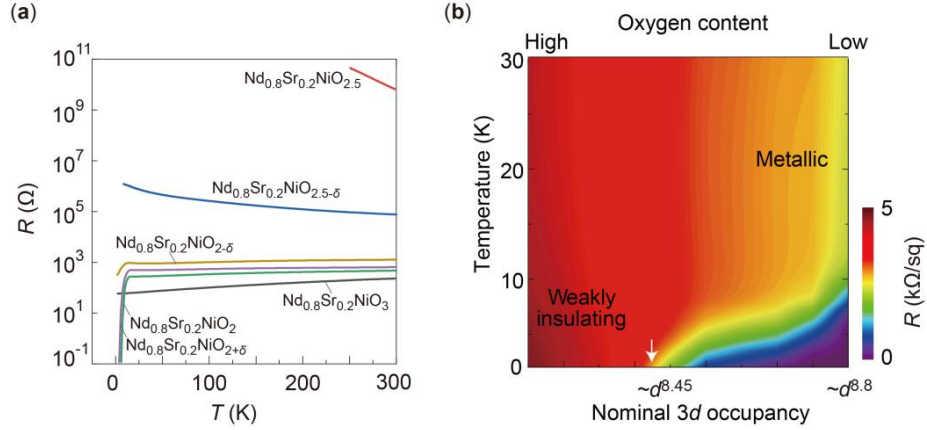

**Fig. 3.** Temperature-dependent resistance of thin films with different oxygen contents. (a) Temperature-dependent resistance of thin films with different oxygen contents corresponding to states in Fig. 2b. The oxygen content is systematically adjusted via sequential reduction cycles. (b) Color-coded phase diagram by fine tuning of the oxygen content near the critical point of SIT. The white arrow marks the quantum critical point during the SIT process. The critical resistance here is about 2.7 kΩ, close to the value $h/(8e^2)$ ~3.2 kΩ.

Electrical transport properties of nickelate thin films are modulated by oxygen content variation through the OMR method. Fig. 3a displays temperature-dependent resistance measurements of thin films with progressively reduced oxygen stoichiometry, systematically controlled through sequential reduction cycles. As oxygen content decreases from $Nd_{0.8}Sr_{0.2}NiO_3$ to $Nd_{0.8}Sr_{0.2}NiO_{2.5}$, the temperature-dependent resistance evolves from metallic behavior across all temperatures to a highly insulating state. Further oxygen depletion reduces the resistance while maintaining insulating characteristics. Notably, superconductivity emerges near an oxygen content of $Nd_{0.8}Sr_{0.2}NiO_2$, but subsequent oxygen depletion degrades superconducting performance. The resistance ratio before and after reduction serves as a key indicator of both metallicity and superconducting performance. The reduction method we have developed is highly reproducible and permits precise control. Through *operando* monitoring, we precisely resolve the insulator-superconductor phase boundary induced by oxygen modulation (Fig. 3b). The white arrow marks the quantum critical point of SIT, where the sheet resistance $R_c$ is about 2.7 kΩ—close to the value $h/(8e^2) \approx 3.2$ kΩ expected for a purely fermionic scenario [46]. Nevertheless, macroscopic inhomogeneity may in principle create metallic shunts that keep the measured sheet resistance below the genuine quantum-critical value; thus, the extracted $R_c$ should be regarded as a lower bound. This approach establishes a viable route for fabricating films with controlled oxygen stoichiometries, enabling systematic investigations into oxygen-dependent superconducting mechanisms.

Ionic liquid (IL) gating has emerged as a prominent technique for modulating the SIT in superconducting systems through carrier density modification or chemical intercalation [47–57]. This method has been particularly effective in oxide superconductors where oxygen and hydrogen intercalation/injection can substantially alter electronic transport properties. However, its application to nickelate superconductors remains relatively unexplored. Previous studies on $Nd_{0.8}Sr_{0.2}NiO_2$ thin films demonstrated material stability under electrostatic gating conditions (e.g. $-1$ to $+4$ V gate voltage, mainly in electrostatic regime) [51]. In this study, we applied an extended gate voltage of $-4$ V at 250 K to explore the electrochemical regime of IL gating in $Nd_{0.8}Sr_{0.2}NiO_2$. Superconductivity was gradually suppressed with increasing gating cycles, as shown by the color-coded resistance evolution in Fig. 4a. Figs. 4b–g present the corresponding temperature-dependent resistance (R-T) curves measured under different applied magnetic fields, revealing a gate-induced SIT at zero magnetic field and a distinct magnetic-field-induced SIT. Post-gating XRD showed a leftward shift of the XRD peaks, indicating a structural change and an increase in oxygen content. Although IL gating induces oxygen re-intercalation together with measurable surface degradation (evidenced by reduced XRD intensities), in contrast to the uniform oxygen removal during vacuum reduction (Fig. 2b), the gated insulating state exhibits Kondo-like $-\ln T$ resistivity identical to that reported for chemically under-doped $Nd_{1-x}Sr_xNiO_2$ films. This indicates that hole doping from oxygen injection, rather than disorder-driven localization, dominates the gating-induced SIT. Thus, IL gating acts as a powerful counterpart to OMR. While the OMR technique enables widespread modulation across the entire phase diagram, IL gating provides an independent validation of the oxygen-driven SIT and offers a valuable method for finer adjustments within the transitional region.

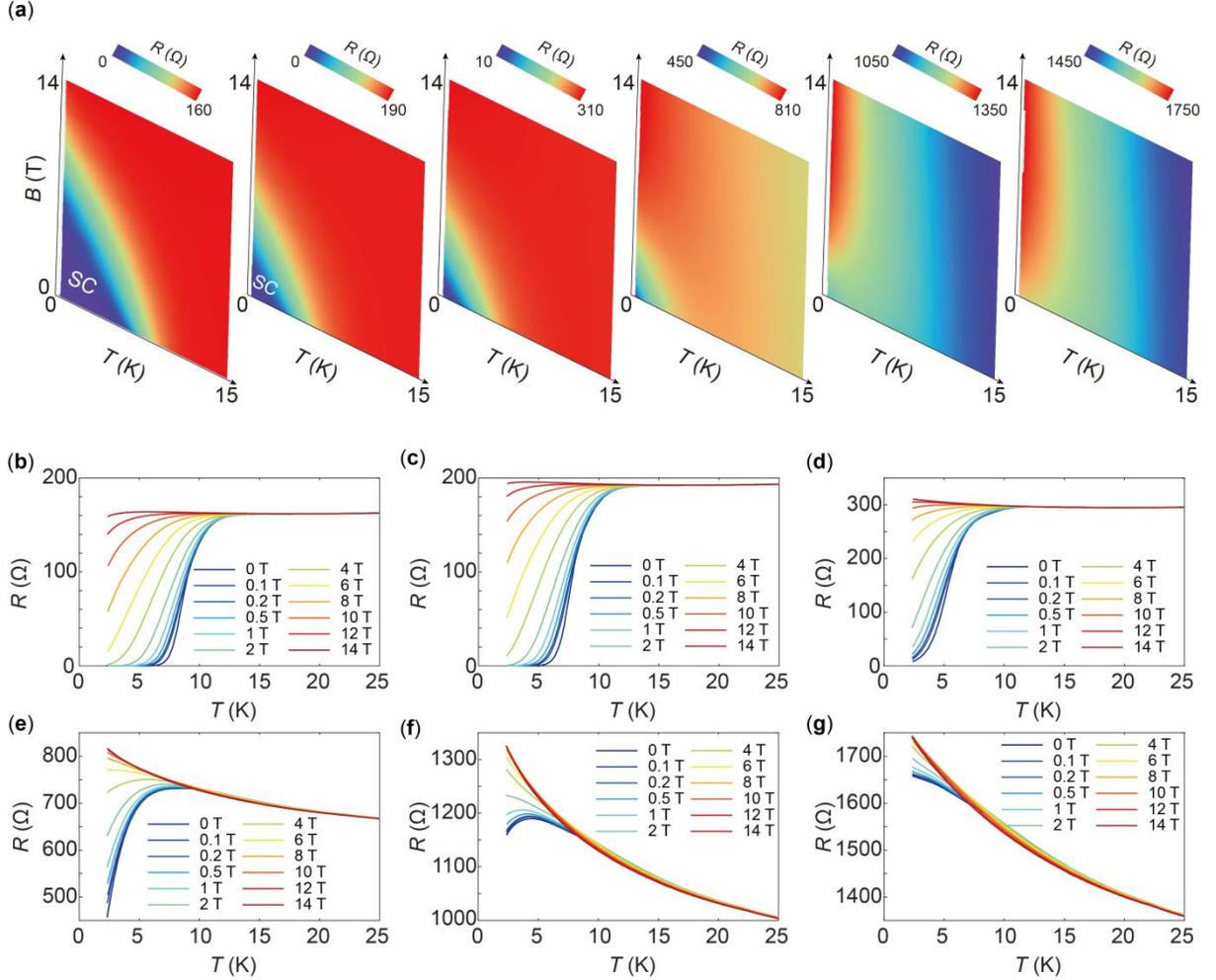

**Fig. 4.** (a) Color-coded phase diagrams of $Nd_{0.8}Sr_{0.2}NiO_2$ as a function of temperature and magnetic field, modulated via IL gating. The IL gating experiment was performed as follows: the sample was held at 250 K under −4 V for 60 mins, then cooled to base temperature while the bias was maintained. Subsequently, still under −4 V, it was reheated to 250 K and held for another 60 mins, during which the resistance rose steadily. Finally, with the voltage still applied, the sample was re-cooled to reach the new gated state. The film undergoes a gradual superconductor to insulator transition. (b–f) *R-T* curves corresponding to panels in (a).

Fig. 5 presents a schematic phase diagram summarizing our findings. An intriguing feature is the finite transitional region, which is characterized by an enhanced Nernst effect indicative of vortex motion. Since the existence of vortices requires the formation of Cooper pairs, the entry into this region from the normal state signifies the onset of pairing. It therefore follows that the oxygen-tuned SIT crossing the transitional region is likely driven by the destruction of the Cooper pairs, rather than the loss of phase coherence while Cooper pairs persist. This complex picture, further distinguished by the mixed 2D/3D nature of the superconductivity, cannot be captured by 2D bosonic models [58]. The insulating ground state does not emerge concurrently with the suppression of superconductivity. The weak insulating regime shows a $-\ln T$ resistivity similar to underdoped $Nd_{1-x}Sr_xNiO_2$ [59], which suggesting Kondo-like scattering as one plausible origin, though Anderson localization or other disorder-driven mechanisms cannot be excluded.

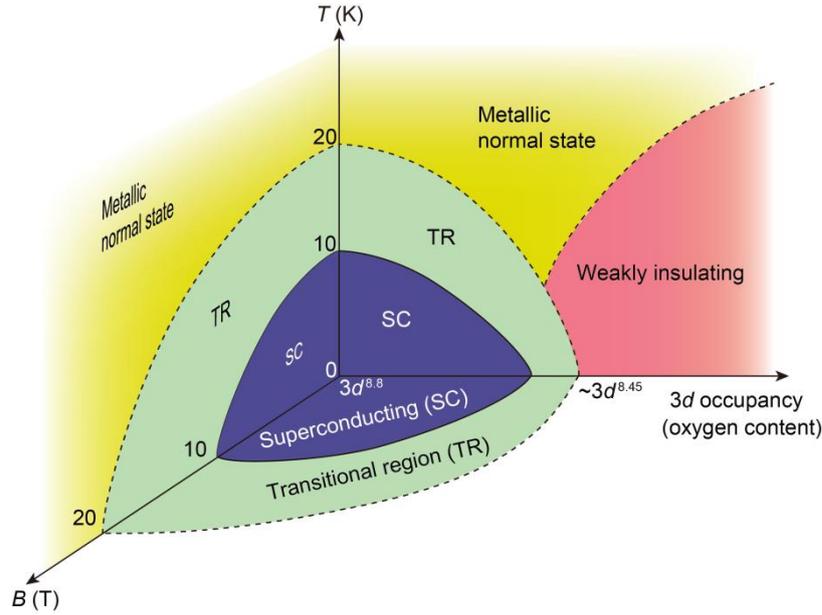

**Fig. 5.** Schematic three-dimensional superconducting phase diagram of the infinite-layer nickelate thin film system with temperature, magnetic field and $3d$ orbital occupancy. This continuous diagram is a schematic guide based on the discrete experimental states shown in Figs. 2–4; it is qualitatively correct but not intended for quantitative accuracy. The "transitional region" color band marks the interval between $T_{c,onset}$ and $T_{c,0}$. The enhancement of Nernst signals is observed exclusively within this band, whereas no vortex-fluctuation signal is detected above $T_{c,onset}$

## 4. Conclusion

In summary, by developing an OMR technique, we have achieved systematic and accurate control across the SIT in infinite-layer nickelates. This capability, combined with STEM, XAS, and multi-parameter transport measurements, enabled us to establish a comprehensive phase diagram in the oxygen content-field-temperature space. Our findings revealed that the structural transition in nickelates is closely linked to the Ni $3d$ orbital occupancy, with variations in oxygen content influencing the electronic structure. Electronically, Nernst effect measurements indicated that pairing in nickelates initiates at the onset of the resistive drop, which is distinct from cuprates, and the subsequent emergence of the Meissner effect at zero resistance marked the establishment of global phase coherence. Angle-dependent magnetotransport measurements suggested a mixture of 2D and 3D superconducting characteristics, indicating that the SIT in nickelates deviates from the canonical two-dimensional model. These results provide valuable insights into the interplay between structural and electronic phase transitions in infinite-layer nickelates and highlight the potential for targeted synthesis and modulation of complex oxide films.

## Conflict of interest

The authors declare that they have no conflict of interest.

## Acknowledgments

We thank Ariando for fruitful discussions. This work was supported by the National Key R&D Program of China (2024YFA1408100 and 2022YFA1403101), the Natural Science Foundation of China (92565303, 92265112, 12374455, 52388201, 12504161, 12504166, 12174325, 12404157, 92365203, U24A6002), the Guangdong Major Project of Basic Research (2025B0303000004), the Guangdong Provincial Quantum Science Strategic Initiative (GDZX2501001 & GDZX2401004 & GDZX2201001), the Shenzhen Science and Technology Program (KQTD20240729102026004), and the Shenzhen Municipal Funding Co-Construction Program Project (SZZX2301004 & SZZX2401001), the China Postdoctoral Science Foundation (2024M761276), the Guangdong Basic and Applied Basic Research Grant (2023A1515011352) and the Research Grants Council (RGC) of the Hong Kong Special Administrative Region, China (CityU 21301221, CityU 11309622, CityU 11300923, CityU 11313325), the Quantum Science and Technology-National Science and Technology Major Project (2021ZD0302502), the Natural Science Foundation of Jiangsu Province (BK20253012, BK20243011, BK20233001), the Open Research Fund Program of Laboratory of Solid State Microstructures, Nanjing University (grant M38040). We acknowledge the support from International Station of Quantum Materials. We thank the staff from BL07U beamline of Shanghai Synchrotron Radiation Facility (SSRF) for assistance of XAS data collection.

## Author contribution

Heng Wang and Xianfeng Wu designed the reduction setups and performed reduction of nickelate films. Heng Wang and Xianfeng Wu performed low-temperature transport measurements. Wei Lv, Zihao Nie, and Yueying Li performed synthesis of nickelate thin films supervised by Guangdi Zhou. Haoliang Huang led the XAS measurements. Haoliang Huang and Xianfeng Wu analyzed STEM data. Cui Ding, Danfeng Li, and all other authors participated discussions. Zhuoyu Chen, Haoliang Huang and Heng Wang wrote the manuscript with input from all other authors. Qi-Kun Xue, Hongtao Yuan, and Zhuoyu Chen supervised the project.